\documentclass[final,3p,times]{elsarticle}
\usepackage{amssymb}
\usepackage{amsmath}

\usepackage{hyperref}
\hypersetup{
    colorlinks=true,
    linkcolor=black,
    citecolor=blue,
    urlcolor=blue
}
\begin{document}

\begin{frontmatter}

\title{Quantum simulation of baryon scattering in SU(2) lattice gauge theory}

\author[a]{Jo\~{a}o Barata}
\author[b]{Juan Hormaza}
\author[c,d,e]{Zhong-Bo Kang}
\author[f,g]{Wenyang Qian\corref{cor1}} 
\cortext[cor1]{Speaker}
\ead{wqian@ccnu.edu.cn}

\affiliation[a]{organization={CERN},
            addressline={Theoretical Physics Department, CH-1211}, 
            city={Geneva},
            postcode={23}, 
            country={Switzerland}}
\affiliation[b]{organization={Universidad Nacional de Colombia},
            city={Manizales},
            country={Colombia}}
            
\affiliation[c]{organization={Department of Physics and Astronomy},
            addressline={University of California,
Los Angeles}, 
            city={Los Angeles},
            postcode={90095}, 
            state={CA},
            country={USA}}
            
\affiliation[d]{organization={Mani L. Bhaumik Institute for Theoretical Physics},
            addressline={University of California,
Los Angeles}, 
            city={Los Angeles},
            postcode={90095}, 
            state={CA},
            country={USA}}
\affiliation[e]{organization={Center for Quantum Science and Engineering},
            addressline={University of California}, 
            city={Los Angeles},
            postcode={90095}, 
            state={CA},
            country={USA}}
\affiliation[f]{organization={Institute of Particle Physics and Key Laboratory of Quark and Lepton Physics (MOE)},
            addressline={Central China Normal University}, 
            city={Wuhan},
            postcode={430079}, 
            state={Hubei},
            country={China}}
\affiliation[g]{organization={Instituto Galego de Fisica de Altas Enerxias (IGFAE)},
            addressline={University of Santiago de Compostela}, 
            city={Santiago de Compostela},
            postcode={15703}, 
            state={A Coruña},
            country={Spain}}

\begin{abstract}
We present a first real-time study of hadronic scattering in a $(1+1)$-dimensional SU(2) lattice gauge theory with fundamental fermions using tensor-network techniques. Working in the gaugeless Hamiltonian formulation, we investigate scattering processes across sectors of fixed global baryon number $B = 0, 1, 2$, corresponding respectively to meson--meson, meson--baryon, and baryon--baryon collisions. At strong coupling, the $B = 0$ and $B = 2$ channels exhibit predominantly elastic dynamics closely resembling the U(1) Schwinger model. The mixed $B = 1$ sector displays qualitatively new behavior: meson and baryon wavepackets become entangled during the collision, with the slower state becoming spatially delocalized while the faster one propagates ballistically. We characterize these processes through local observables, entanglement entropy, and the information lattice.
\end{abstract}
 
 
\end{frontmatter}
 
\section{Introduction}
\label{sec1}
 
Understanding the real-time, non-perturbative dynamics of hadronic scattering in gauge theories remains one of the central open challenges of quantum field theory. Traditional approaches---perturbation theory and Euclidean lattice QCD---are ill-suited for these questions, due respectively to the breakdown of perturbative series at strong coupling and the sign problem obstructing real-time continuation~\cite{Asakawa:2000tr}. New theoretical tools rooted in quantum information science (QIS), including tensor network methods, quantum computers, and analog quantum simulators, have opened powerful avenues toward this goal~\cite{Zohar:2021nyc}.
 
Among these, tensor-network methods based on matrix product states (MPS) have proven particularly effective in $(1+1)$ dimensions, providing high-precision access to the non-equilibrium dynamics of strongly coupled gauge theories~\cite{Banuls:2018jag}. Recent work has achieved first simulations of elastic and inelastic scattering in Abelian (U(1)) gauge theories~\cite{Papaefstathiou:2024zsu}. However, extending this program to non-Abelian models is a crucial step toward capturing genuinely QCD-like phenomena such as confinement, baryon formation, and flavor dynamics.
 
In this proceedings we highlight our recent work~\cite{Barata:2025rjb} on real-time hadronic scattering in a $(1+1)$D SU(2) lattice gauge theory. The key new feature compared to Abelian models is the conserved baryon number $B$, which allows for the formation of diquark baryon states and richer hadronic spectra. We study scattering processes for global $B=0,1,2$ --- meson--meson, meson--baryon, and baryon--baryon collisions --- using tensor network.
 
\section{SU(2) gauge theory in (1+1) dimensions}
\label{sec2}

We consider SU(2) gauge theory in $(1+1)$ dimensions coupled to fundamental fermions, which is the simplest non-Abelian generalization of the Schwinger model sharing key features such as confining potentials~\cite{Coleman:1976uz}. The essential new ingredient relative to the U(1) case is a global baryon-number symmetry, which supports gauge-invariant color-singlet states with unequal numbers of quarks and antiquarks, giving away rich structures and dynamics.
 
\textbf{Lattice Hamiltonian: }Starting from the continuum Hamiltonian in temporal gauge $A_0^a=0$, we adopt the Kogut--Susskind staggered-fermion prescription~\cite{Kogut:1974ag} with lattice spacing $a$ and $N$ sites. The resulting lattice Hamiltonian
\begin{align}
    {H}_l = \frac{1}{2a} \sum_{n=1}^{N-1} \big( {\phi}^\dagger_n \, {U}_n \, {\phi}_{n+1}+ \text{h.c.} \big) + m \sum_{n=1}^{N} (-1)^n {\phi}^\dagger_n {\phi}_n + \frac{a g^2}{2} \sum_{n=1}^{N-1} {L}^2_n \,,
\end{align}
where $\phi_n = (\phi_n^1, \phi_n^2)^T$ is a two-component (colored) staggered spinor, $U_n \in \mathrm{SU}(2)$ is the gauge link, and $L_n^a$ is the electric field operator. Just like the U(1) case, Gauss's law constrains physical states: $G^a_n|\text{phys}\rangle = 0$, where $G^a_n = L^a_n - R^a_{n-1} - Q^a_n$ and $Q^a_n = \phi_n^\dagger T^a \phi_n$ is the color charge operator at site $n$. A key simplification of $(1+1)$D gauge theories is that the gauge field acts as a non-dynamical potential and can be explicitly integrated out. Using the residual gauge freedom, one performs a site-dependent color rotation which removes the gauge links from the kinetic term, yielding a purely fermionic Hamiltonian with long-range interactions mediated by the electric flux.
Notably, this \emph{gaugeless} formulation requires no truncation of the electric flux.
 
\textbf{Spin Hamiltonian: }Since there are only fermionic degrees of freedom in the Hamiltonian, a Jordan-Wigner transformation, i.e, ${\psi}_n \equiv \prod_{k < n} (-\sigma^z_k)\,  {\sigma}_n^{-}$, can be used to map the two components $(\phi_n^1,  \phi_n^2)$ of the fermions at site $n$ to $(\psi_{2n-1},  \psi_{2n})$ on a spin-$\tfrac{1}{2}$ chain. 
 The resulting spin Hamiltonian contains the kinetic energy term, the mass terms, and long-range color-exchange interactions $H_E$: 
\begin{align}
    H_s &=  -\frac{1}{2a} \sum_{n=1}^{N-1} \left[ \sigma_{2n}^+ \sigma^z_{2n} \sigma_{2n+1}^- + \sigma_{2n+1}^+ \sigma^z_{2n+1} \sigma_{2n+2}^- + \text{h.c.} \right] 
    + ma \sum_{n=1}^N \left( \frac{(-1)^n}{2} ( \sigma^z_{2n-1} + \sigma^z_{2n} )  \right) + H_E\, ,
\end{align}
and the physical content at each qubit site is determined by its parity $n \bmod 4$: sites with $n\equiv 1,2$ encode antiquarks ($\bar{r}$ and $\bar{g}$ respective), while $n\equiv 3,0$ encode quarks ($r$ and $g$ respectively). Now the physical problem is discretized and converted into spin Hamiltonian that can be simulated with tools like quantum computer and tensor network. In this work, we use tensor network simulation with \texttt{iTensor} library~\cite{itensor} to obtain our numerical results.

\section{Real-time simulation of hadronic scattering}
\label{sec3}

The main idea of the simulation is the follow: Two wavepackets with respective baryon number $B$ and opposite momenta are initialized on opposing sides of the chain and evolved under $H_s$ Hamiltonian. 

\textbf{Preparation of incoming wavepackets: }As a first step, we prepare hadronic wavepackets as Gaussian-modulated superpositions of local hadron creation operators acting on the vacuum $|\Omega\rangle$:
\begin{align}
    |\psi_B(k;\sigma,c)\rangle \propto \sum_{n} e^{-(n-c)^2/(2\sigma^2)}\, e^{ink}\, \mathcal{A}_B(n)\,|\Omega\rangle\,,
\end{align}
where $\mathcal{A}_B(n)$ is the local hadron creation operator at site $n$ for baryon number $B$. For mesons ($B=0$) it involves nearest-neighbor quark--antiquark pairs; for baryons ($B=\pm 1$) it creates a local diquark at the appropriate staggered sites. The vacuum $|\Omega\rangle$ is obtained as the ground state to our Hamiltonian using DMRG algorithm~\cite{schollwock2011density} and aftewards the wavepacket is constructed by contracting local tensor operators. 
The main simulation results are performed at $ga=5$, $ma=0.2$ ($\mu=2m/(ag^2)=0.016$, $x=1/(ag)^2=0.04$) on $N=60$ qubits. DMRG is run with $N_\text{sweep}=2000$ sweeps at maximum bond dimension for the MPS equal to $\chi_\text{max}=80$  (with convergence checks up to $\chi_\text{max}=200$) at a cutoff $\epsilon=10^{-12}$. The subsequent real-time evolution uses the TDVP algorithm~\cite{Haegeman:2011zz} with $\chi_\text{max}=80$ and time step $a\Delta t=0.1$. In particular, We study three sectors, depending on the total baryon number $B$ of the setup.

\textbf{$B=0$, meson--meson case.} Symmetric scattering with equal maximal momenta is perfectly elastic: the two mesons pass through each other with unchanged structure, and the entanglement entropy returns to its initial profile after the collision. For asymmetric momenta, the process remains elastic.
 
\textbf{$B=1$, meson--baryon case.} This sector reveals qualitatively new physics. The baryon remains approximately localized while the meson partially reflects and partially tunnels. After the collision, the meson probability density is spread across the lattice, suggesting two wavepackets form a single entangled collective state than separating cleanly.
 
\textbf{$B=2$, baryon--baryon case.} Two baryons scatter elastically, closely resembling the U(1) Schwinger model dynamics. The kinematic access to inelastic channels is blocked because the maximum baryon momentum lies far below the momentum threshold for producing two mesons from $B+\bar{B}$.

Local observables are extensively studied, such as the baryon number operator $B_n = \sigma_n^z/4$ and chiral condensate $C_n = \bar\psi_n\psi_n$ are directly employed for studying internal color structure of the colliding states. The chromoelectric field and energy $E_n \equiv \langle L_n^2 \rangle = \langle (\sum_{m\leq n} Q_m )^2 \rangle$ for the gauge field at each link can also be reconstructed from the color charges of the sites to its left to track the color strings. See Ref.~\cite{Barata:2025rjb} for more details.
\begin{figure}[t]
\centering
\includegraphics[width=0.80\textwidth]{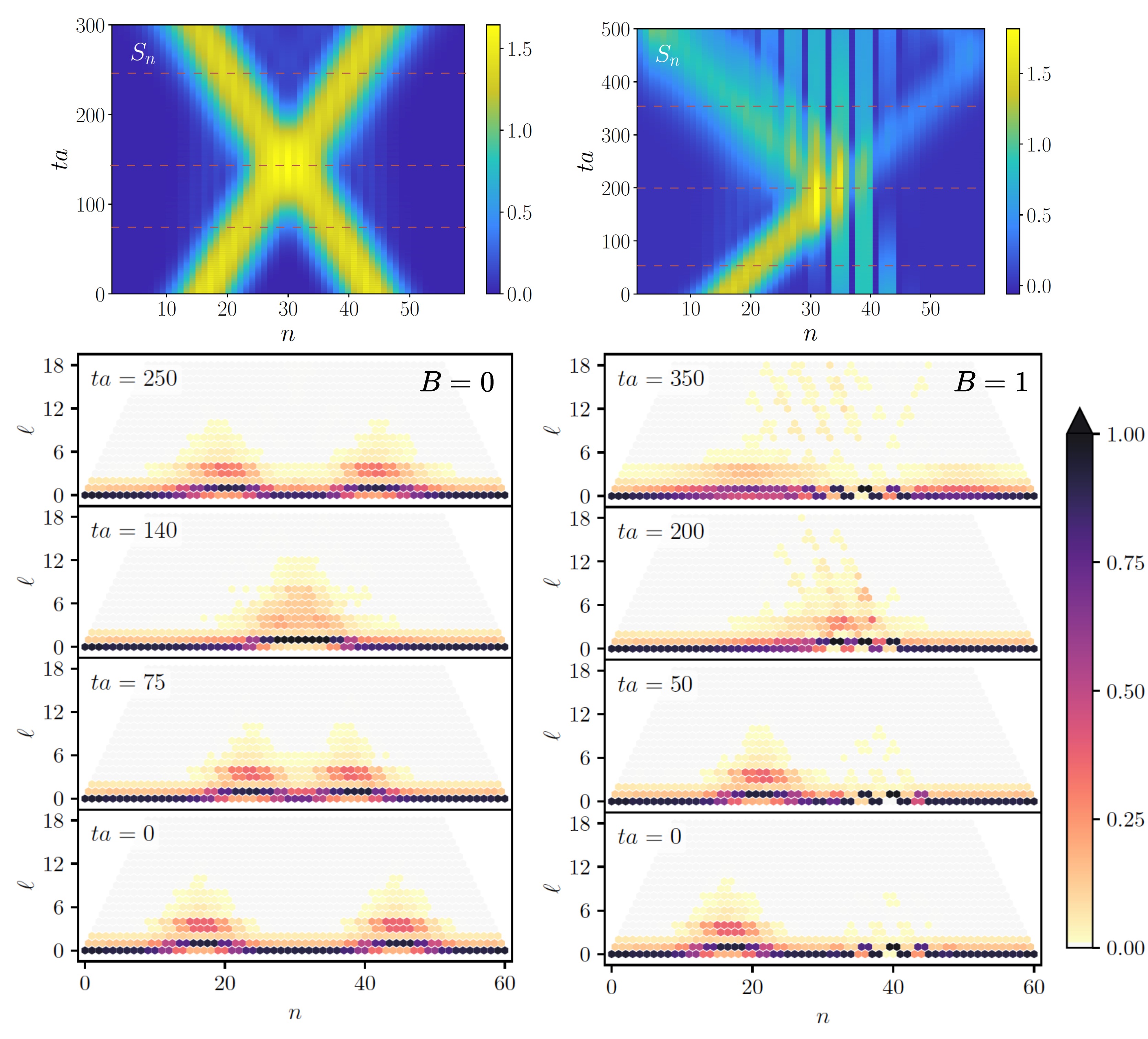}
\caption{Entanglement entropy and lattice information for $B=0$ meson-meson (\textbf{left}) and $B=1$ meson-baryon scattering (\textbf{right}).}\label{fig-1}
\end{figure}

\textbf{Novel observable: }Beyond entanglement entropy, we employ the \emph{information lattice}~\cite{Kvorning:2021rdf} as a new diagnostic of correlations. For each subsystem block of $\ell$ consecutive sites at position $n$, we define the local information
\begin{align}
    i(n,\ell) &= I(\rho^\ell_n) - I(\rho^{\ell-1}_{n-1/2}) - I(\rho^{\ell-1}_{n+1/2}) + I(\rho^{\ell-2}_n)\,, 
\end{align}
where $I(\rho) = \log_2[\dim(\rho)] - S(\rho)\,$ and $S(\rho)$ is the von Neumann entropy. Non-zero $i(n,\ell)$ signals genuine $\ell$-body correlations not reducible to smaller blocks.
In Fig.~\ref{fig-1}, we present both entanglement entropy $S_n$ and lattice information $i(n,l)$ for $B=0,1$ scattering. In the $B=0$ elastic case, the information lattice shows a characteristic peak at $\ell\approx 3$--$4$ for each wavepacket, which is transiently enhanced during overlap and restored afterward. In the $B=1$ sector, no new length scales $\ell$ are populated significantly, indicating that the collective entangled state does not form a bound state with a distinct internal correlation structure. The lattice information complements the entanglement entropy, providing a spatial correlation picture. 
 
\section{Summary and Conclusions}
\label{sec:sumandconc}
 
We have presented a first real-time study of hadronic scattering in a $(1+1)$D SU(2) lattice gauge theory using tensor network methods in the gaugeless Hamiltonian formulation. We simulated meson--meson ($B=0$), meson--baryon ($B=1$), and baryon--baryon ($B=2$) collisions at strong coupling, where the $B=0$ and $B=2$ sectors exhibit predominantly elastic dynamics qualitatively similar to the Schwinger model; the $B=1$ sector reveals genuinely non-Abelian physics: the meson and baryon wavepackets become entangled and form a collective state. Our results demonstrating the utility of QIS-inspired diagnostics for probing hadronic dynamics. 
 
\section*{Acknowledgments}
 
We are grateful to Mari Carmen Ba\~nuls, Meijian Li, and Enrique Rico for helpful discussions and collaboration. 
ZK is supported by NSF under grant No.~PHY-2515057. 
WQ is supported by the ERC under project ERC-2018-ADG-835105 YoctoLHC; by Maria de Maeztu excellence unit grant CEX2023-001318-M and project PID2020-119632GB-I00; by ERDF/EU; by the MSCA Fellowships under grant No.~101109293; and by Xunta de Galicia.

\end{document}